\documentclass[12pt,a4paper]{article}
\usepackage{amsmath,latexsym,amsfonts,amssymb}
\usepackage{graphicx}
\usepackage[super,numbers]{natbib}

\newtheorem{example}{Example}

\begin{document}
\date{\today}
\setcounter{page}{1}

\title{Shielding the vulnerable in an epidemic:\\ a numerical approach}
\author{Guus Balkema\\
University of Amsterdam}
\maketitle

\begin{abstract} The death toll for Covid-19 may be reduced by dividing the population into two classes, the vulnerable and the fit, with different lockdown regimes.  Instead of one reproduction number there now are four parameters. These make it possible to quantify the effect of the social distancing measures. There is a simple stochastic model for epidemics in a two type population. Apart from the size of the population of the vulnerable and the fit, and the initial number of infected in the two classes, only the four reproduction parameters are needed to run the two type Reed-Frost model. The program is simple and fast. On a pc it takes less than five minutes to do a hundred thousand simulations of the epidemic for a population of the size of the US. Epidemics are non-linear processes. Results may be counterintuitive. The average number of vulnerable persons infected by an infectious fit person is a crucial parameter of the epidemic in the two type population. Intuitively this parameter should be small. However simulations show that even if this parameter is small the death toll may be higher than without shielding. Under certain conditions increasing the value of the parameter may reduce the death toll. The article addresses these blind spots in our intuition. 
\end{abstract}

\setcounter{section}{0}
\setcounter{equation}{0}

\section{Introduction}\label{s0}

The reproduction number for Covid-19 lies between 2 and 3 if we take no action to stop it spreading. With proper measures such as social distancing it may be reduced to below one. It is known that infections are more lethal for the elderly (and for persons who suffer from obesity, diabetes, high blood pressure) than for healthy young persons~\cite{LCW20}. These two facts suggest that a social distancing policy which takes into account the difference in risk for the vulnerable and for the fit might be effective in reducing the overall mortality. So consider a population consisting of a million vulnerable persons and two million fit persons. Assume that the mortality is ten times as high for the vulnerable as for the fit. To be concrete, assume an Infection Fatality Rate (IFR) of 0.01 for the vulnerable and 0.001 for the fit. This yields an overall IFR of $0.004=(0.01+2*0.001)/3$. 

Assume a two type model. A vulnerable person infects $r_v=0.7$ persons on average, a fit person $r_f=1.3$. Ten vulnerable persons and twenty fit, all infectious, on average will yield 7 + 26 = 33 new infections (corresponding to the overall reproduction number $r_0=1.1$). Start with a hundred infections among the vulnerable and two hundred among the fit. The population is compartmentalized. If the compartments are watertight, there is no contact between the vulnerable and the fit. The epidemic will die out among the vulnerable since $r_v=0.7<1$, but it will be more severe among the fit than in the corresponding homogeneous model with reproduction number $r_0=1.1$ since $r_f=1.3>1.1$. Computations show that the total number of deaths is lower than in a homogeneous population.  

In a more realistic model the expected number $r_{f,v}$ of vulnerable persons infected by an infectious fit person and the number $r_{v,f}$ of fit persons infected by an infectious vulnerable person are positive. So suppose ten infectious fit persons infect on average eleven fit persons and two vulnerable persons, and ten infectious vulnerable persons infect five vulnerable persons and two fit persons. This ``standard model'' will play a prominent role in the discussion below. The infection rate between the two groups is low.  Such a society may be said to shield the vulnerable. How effective is this shield? Simulations show that in the heterogeneous society there will be more deaths than in a homogeneous population with reproduction number $r_0=1.1$. In the homogeneous population the death toll has a mean value of 2124; in the two type model the mean is ten per cent higher, 2347. 

The result is not implausible.  There is a considerable increase in the total number of infections among the fit.  The positive cross infection $r_{f,v}=0.2$ from the fit to the vulnerable has the effect that the vulnerable are pulled along in this more severe epidemic, yielding a higher overall mortality.  Compartmentalization increases the death toll.

Shielding the vulnerable  may be counterproductive. That is what this paper is meant to show.  For three values of $r_0$, 1.1, 1.0 and 0.95, we choose values of $r_v$ and $r_f$ which satisfy $(r_v+2r_f)/3=r_0$. We then plot the mortality as the transition rate $r_{f,v}$ from fit to vulnerable varies between 0 and 0.8. This is done for various values of $r_{v,f}$. Figures~\ref{fepi1}, left side, and~\ref{fepi2} show the effect on the mortality. 

It is not clear how the transitions $r_{f,v}$ and $r_{v,f}$ are related. Intuitively for nursing homes one might interpret $r_{f,v}$ as measuring infections caused among the elderly by visits by the family, and $r_{v,f}$ as infections among the nurses and staff caused by illness among the elderly. Section~\ref{s8} will look at this issue more closely. 

The main purpose of the paper is to exhibit possible adverse effects of shielding in a heterogeneous population. This aim has been achieved already above by mentioning the results of the computations for the standard model. For the given values $r_v=0.7$ and $r_f=1.3$ above, and $r_{v,f}=0.2$, a large value of $r_{f,v}$ will \emph{reduce} the total mortality. Based on that result the government might consider launching a campaign: ``Tonight don't meet at the pub; visit your granny instead."

The benefits of social distancing measures which differentiate between the fit and the vulnerable are incontestable. This paper shows that policies have to be chosen with care to avoid adverse affects. Here we should mention a different beneficial effect of variations in the reproductive number across a population. Heterogeneity may reduce the herd immunity threshold~\cite{B20, G20}.

For an introduction to the mathematical background see~\cite{B10, B19}. The exposition below is self-contained. It relies on simulations. Readers with some experience in {\tt R}\  are invited to use their skill to explore the effects of variations in the reproduction matrix.  

Section~\ref{sm} introduces the binomial Reed-Frost model. Section~\ref{sr} presents the results. These are discussed in Section~\ref{s8}. Section~\ref{s9} contains our conclusion. The Appendix contains two sections: a non technical explanation of the decrease in mortality associated with large values of $r_{f,v}$ and a discussion of the role of eigenvalues and eigenvectors in the two-type Reed-Frost model.

\section{The model and its program}\label{sm}

In a homogeneous population of size $n_0$ with initially $i_0$ infections the probability of non-infection for any susceptible member of the population is $q=q_0^{i_0}$ where $q_0=1-p_0$ and $p_0=r_0/n_0$ is the probability of infection in a homogeneous population of size $n_0$ with reproduction number $r_0$.  The total number of new infections $i$ among the $n=n_0-i_0$ susceptibles is binomial-$(n,p)$ with $p=1-q$. This yields a recursion  starting with $n=n_0, i=i_0$. In {\tt R}\  the recursion consists of three commands:
\begin{verbatim}
   n<- n-i;  p<- 1-q0^i;  i<- rbinom(1,n,p);
\end{verbatim}
The sequence of commands runs while $i$ is positive. The total number infected is $j=n_0-n$. This is the \emph{binomial} or \emph{Reed-Frost} model. In her exposition of this model in 1952 Abbey~\cite{A52} writes: ``Epidemics can be calculated stepwise from this model, with the aid of random numbers and a table of the cumulative binomial distribution (National Bureau of Standards, 1949)"  and then explains how to use the seven digit tables of random numbers to create realizations of binomial variables and by repetition realizations of the epidemic, closing with the advice: ``For values beyond the range of the binomial tables, the Poisson or normal distributions may be used as approximations to the binomial." Seventy years later for population size $n_0=3*10^6$ a batch of a hundred thousand simulations takes 72 seconds on a ten year old iMac OS 10.11.6 with a 3.06 GHz Intel Core processor  and 1067 MHz DDR3 memory modules in two out of four memory slots. The program determines the mean and sd of $j$, the total number infected. If the initial number of infections is small the epidemic may die out, but for $i_0=300$ initial infections each of the $10^5$ simulations gives rise to a full blown epidemic. 

The program for the multitype binomial model is  similar. Start with a population $n_0=(10^6, 2*10^6)$ of a million vulnerable and two million fit persons and $i_0=(100,200)$ initial infections, a hundred among the vulnerable and two hundred among the fit. The probability that the infection will die out may be neglected. The reproduction number $r_0$ now becomes a matrix $R$ which is akin to the transition matrix in a Markov chain where the $k$th row contains the probabilities $p_{k,m}$ of a transition from state $k$ to $m$. We first look at a specific case, the standard model discussed in the introduction:
\begin{equation}\label{qmR}
R=\begin{pmatrix} r_{v,v}&r_{v,f}\\ r_{f,v}&r_{f,f}\end{pmatrix}=
\begin{pmatrix} 0.5&0.2\\
0.2&1.1\end{pmatrix}.
\end{equation}\noindent
The first row states that ten infectious vulnerable persons will infect on average five vulnerable and two fit persons. The second row states that similarly ten infectious fit persons will infect on average two vulnerable and eleven fit persons. 

\begin{figure}[htp]
\centering
\includegraphics[width=0.48\linewidth]{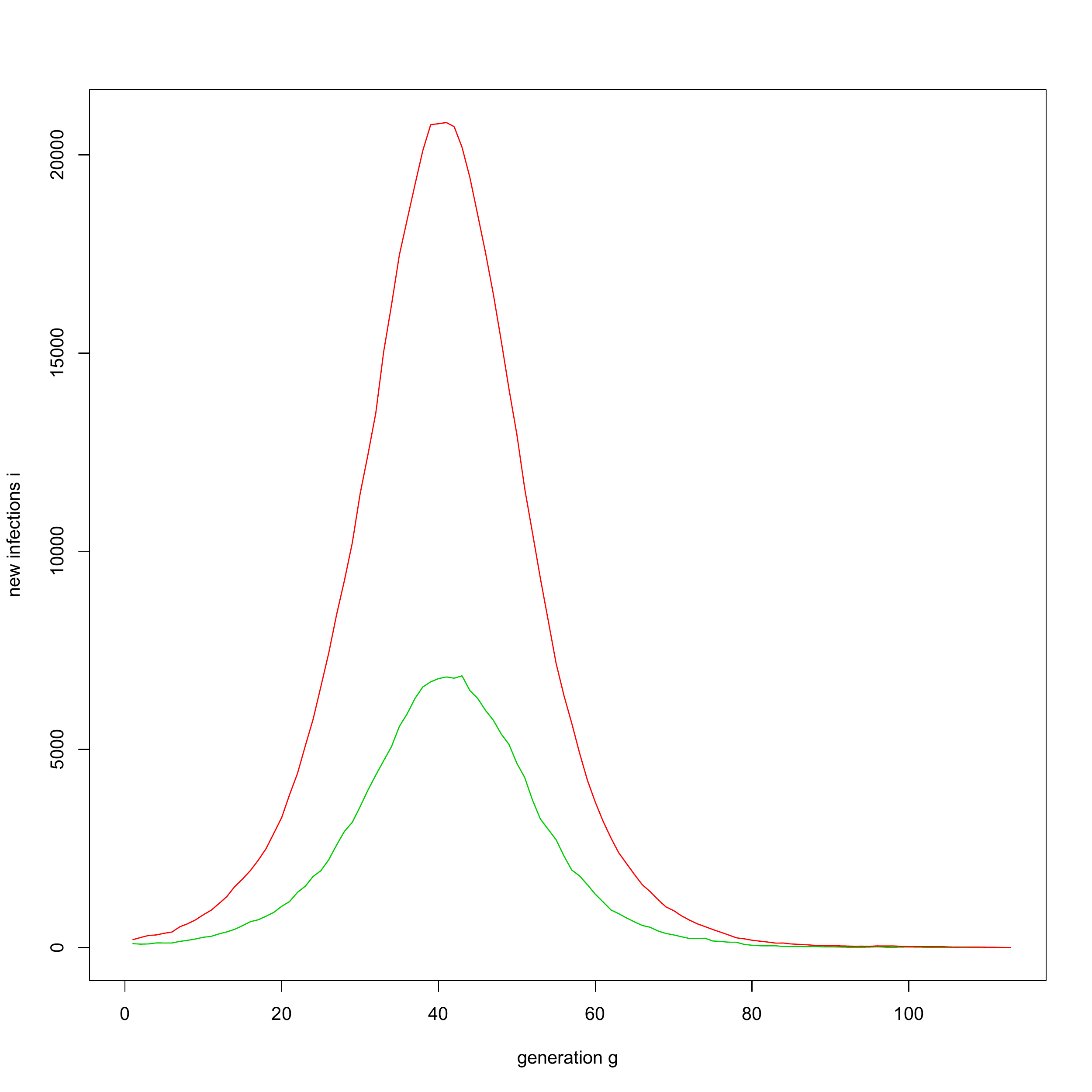}
\hspace{0pt}
\includegraphics[width=0.48\linewidth]{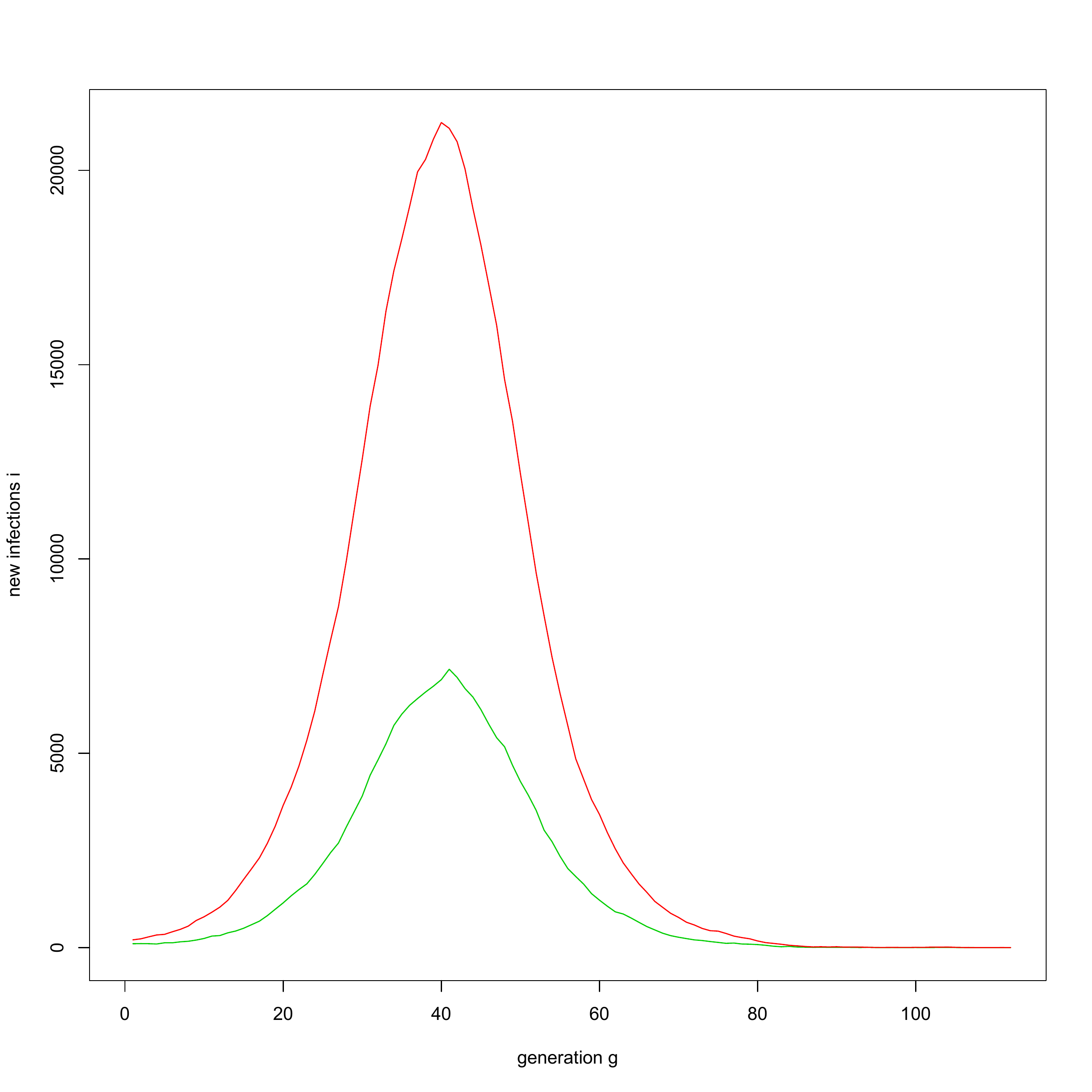}
\caption{Two simulations of the epidemic with the reproduction matrix $R$ in~(\ref{qmR}). The green curve describes the epidemic for the vulnerable, the red curve for the fit.} 
\label{fepi} 
\end{figure}

For a vulnerable person the probability of not being infected is 
$q_v=(1-r_{v,v}/n_v)^{i_v}*(1-r_{f,v}/n_v)^{i_f}$.
A similar expression hold for a fit person. The counterpart of the constant $q_0=1-r_0/n_0$ in the homogeneous model is the matrix:
\begin{equation}\label{qmQ}
Q=
\begin{pmatrix} 1-R[1,1]/n_0[1]&1-R[1,2]/n_0[2]\\
1-R[2,1]/n_0[1]&1-R[2,2]/n_0[2]\end{pmatrix}.
\end{equation}\noindent
The program for a simulation of the epidemic in the multitype model then is:
\begin{verbatim}
   n<- n0; i<- i0;
   while(max(i)>0){
      n<- n-i;
      q1<- Q[1,1]^i[1]*Q[2,1]^i[2]; q2<- Q[1,2]^i[1]*Q[2,2]^i[2];
      i[1]<- rbinom(1,n[1],1-q1); i[2]<- rbinom(1,n[2],1-q2)}
   j<- n0-n;
\end{verbatim}
\begin{equation}\label{qP}
{\rm The\ {\tt R}\ program\ for\ the\ two\ type\ binomial\ Reed-Frost\ model.}
\end{equation}\noindent
List the successive pairs $i$, starting with $i_0=(100,200)$, and plot the infections $i_k[v]$ and $i_k[f]$ to obtain logistic curves as in Figure~\ref{fepi}.

For a batch of a hundred thousand simulations of the epidemic we compute the mean $\mu$ of $j=n_0-n$, the total number of infections, the covariance matrix, the  sd $\sigma$ of the components $j_v$ and $j_f$ and the correlation $\rho$:
$$\mu=(180 387, 543 119)\qquad \sigma=(1362,3988)\qquad\rho=0.88.$$
The mean number of deaths is $1803.9+543.1=2347.0$. The sd of the mean is 0.054.

Compare this to the situation of complete mixing. The reproduction matrix 
\begin{equation}\label{qR11}
R=
\begin{pmatrix} 11/30&22/30\\
11/30&22/30\end{pmatrix}
\end{equation}\noindent
yields:
$$\mu=(176990,353978)\qquad\sigma=(2341,4651)\qquad\rho=0.980\qquad d=2123.9.$$

\section{Results}\label{sr}

\begin{figure}[htp]
\centering
\includegraphics[width=0.48\linewidth]{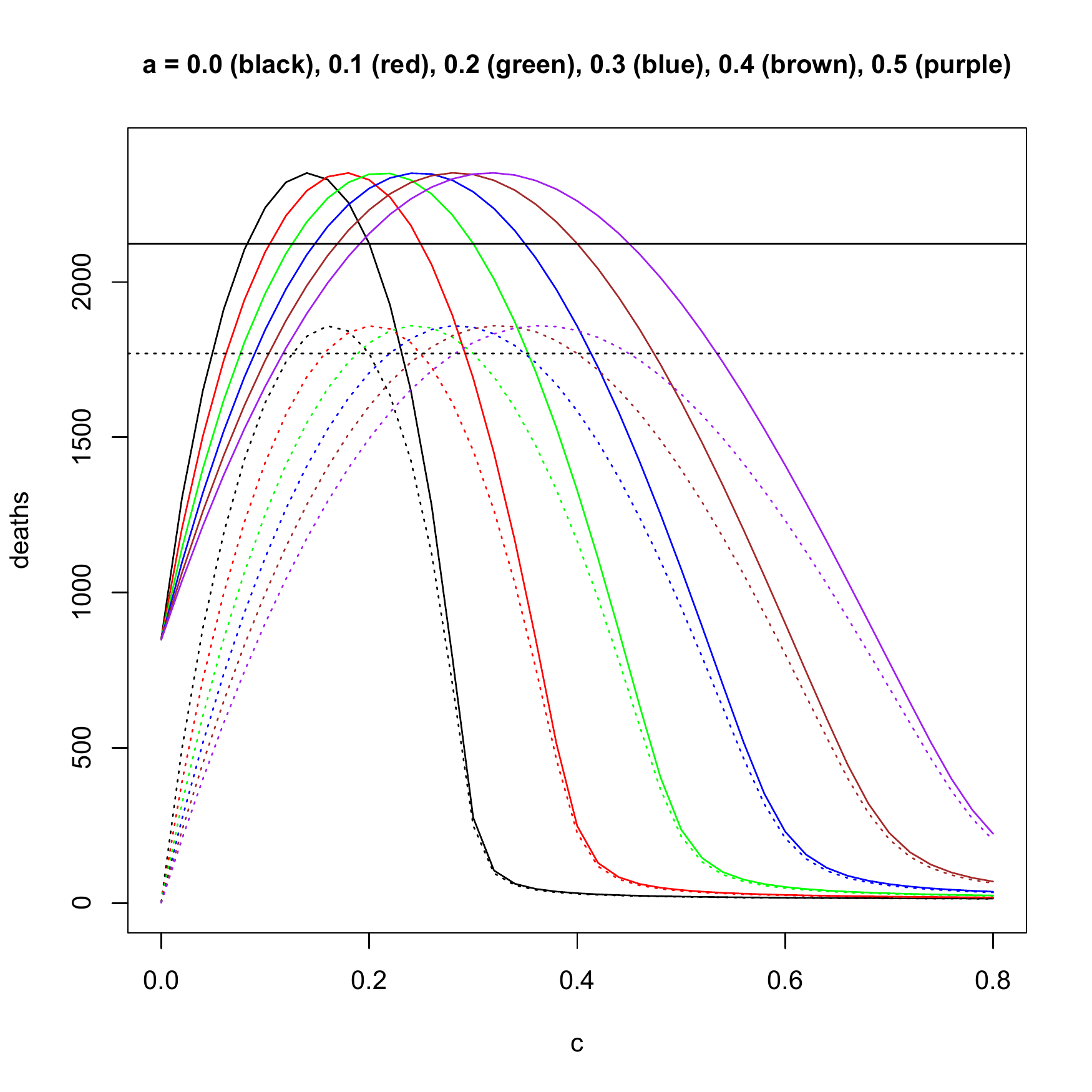}
\hspace{0pt}
\includegraphics[width=0.48\linewidth]{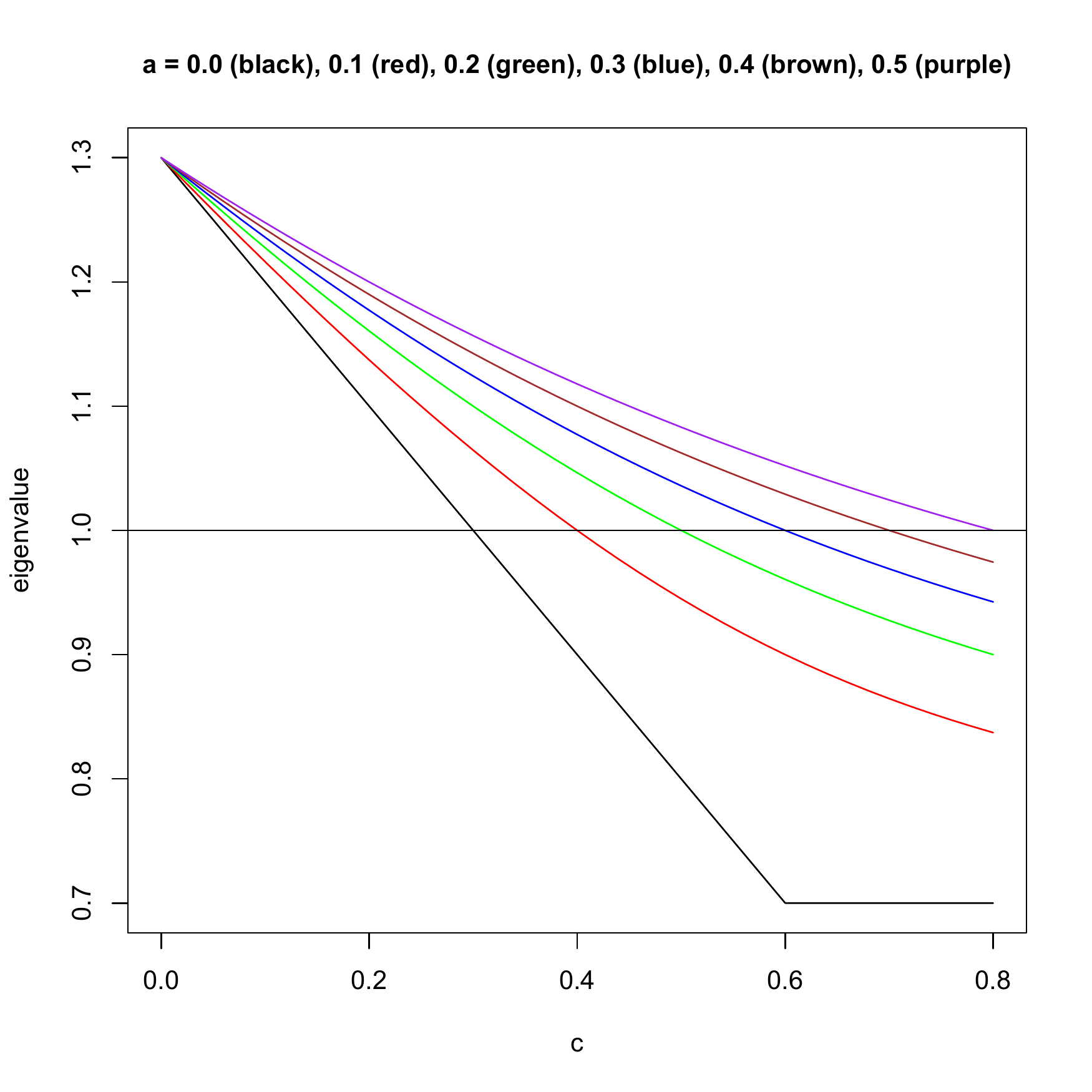}
\caption{The two type model with reproduction matrix $R(a,c)$ in~(\ref{qR}) for $0\le c\le 0.8$ and $a=0, 0.1,\ldots,0.5$.
On the left the mortality; on the right the maximal eigenvalue.} 
\label{fepi1} 
\end{figure}

\begin{figure}[htp]
\centering
\includegraphics[width=0.48\linewidth]{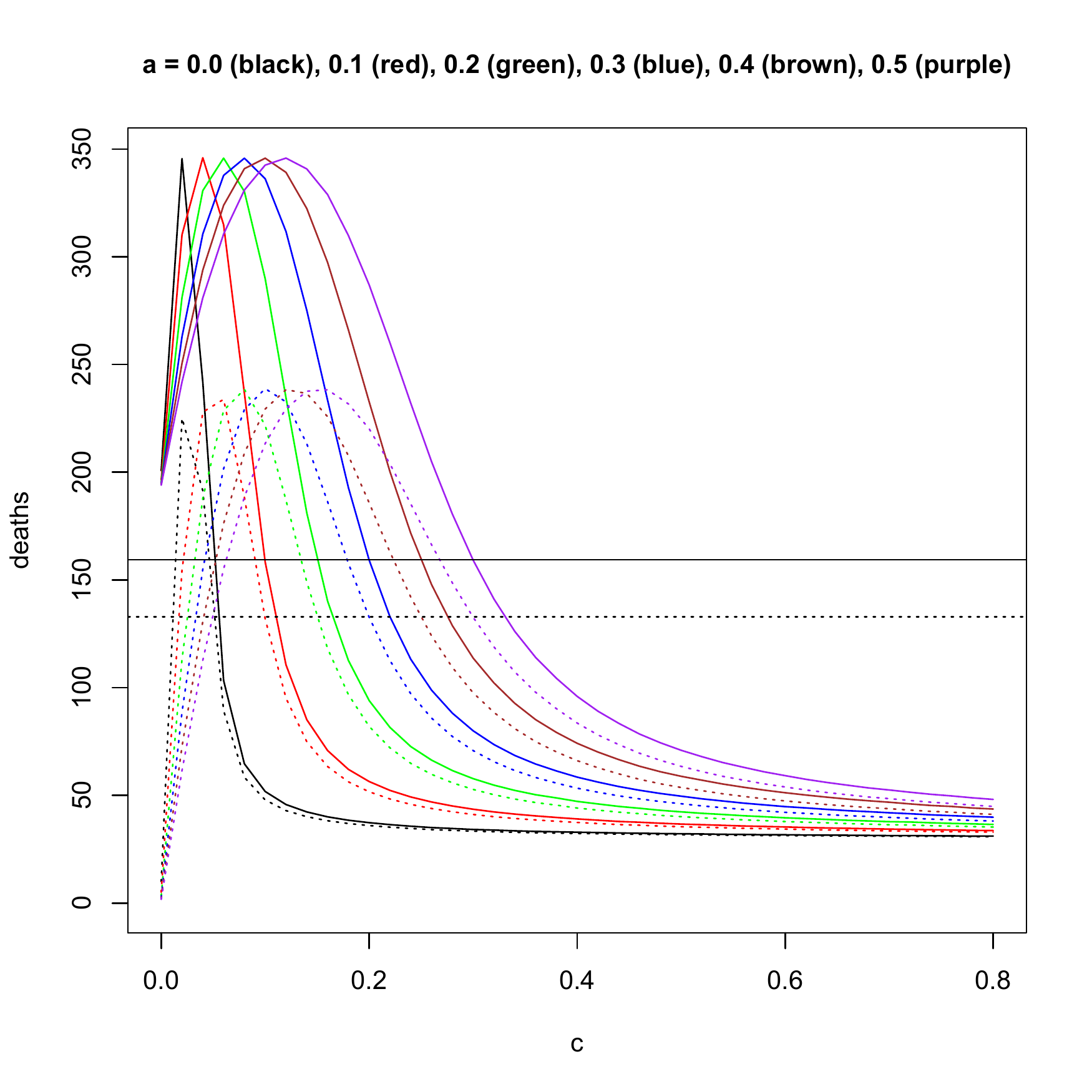}
\hspace{0pt}
\includegraphics[width=0.48\linewidth]{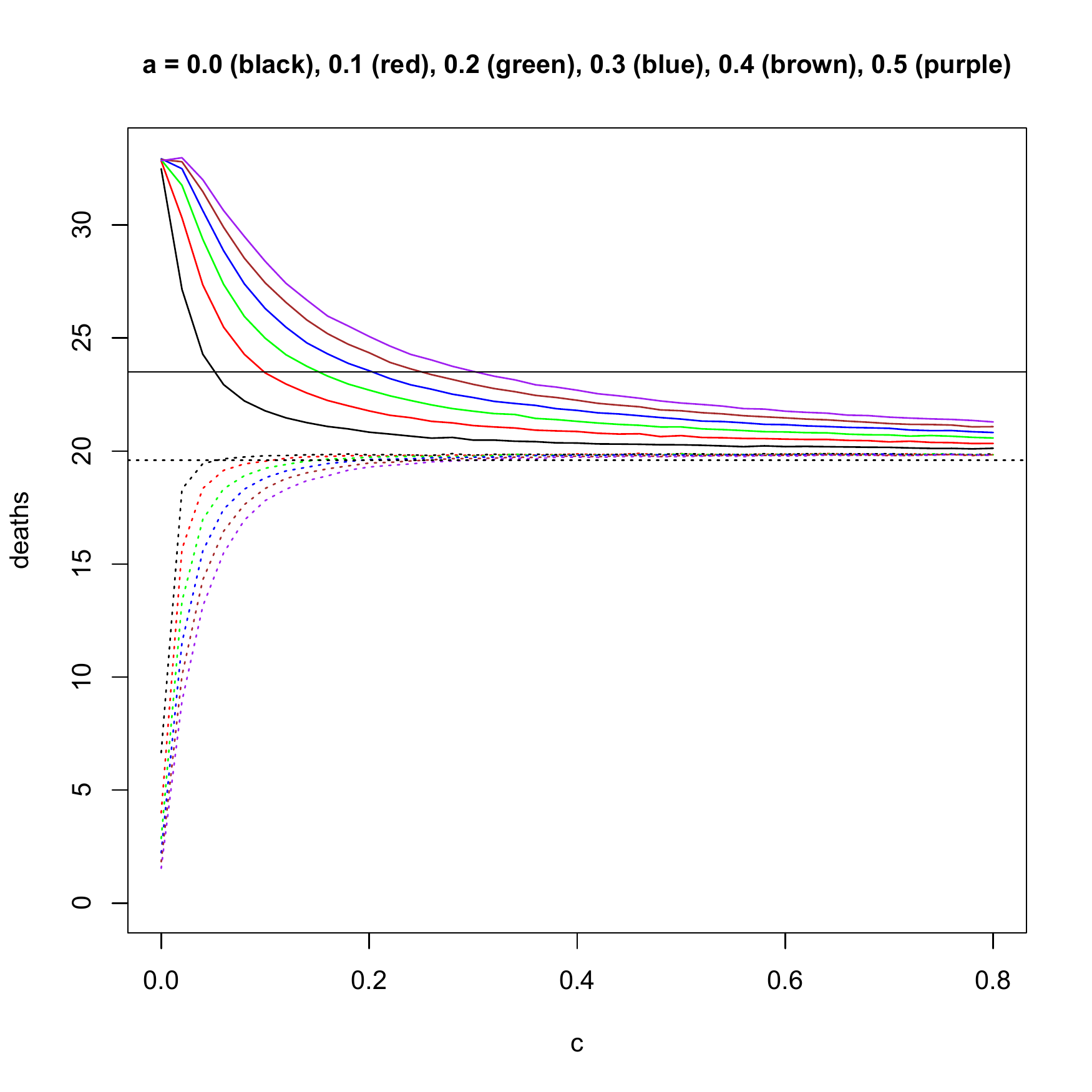}
\caption{The mortality for $R_{a,c}$ with $0\le c\le 0.8$ and $a$ in $0:5/10$. On the left $r_v=0.9$, $r_f=1.05$ and $r_0=1$. On the right $r_v=0.85$, $r_f=1.0$ and $r_0=0.95$. } 
\label{fepi2} 
\end{figure}

With $r_v=0.7$ and $r_f=1.3$ the reproduction matrix has the form
\begin{equation}\label{qR}
R=R(a,c)=\begin{pmatrix} 0.7-a&a\\ c&1.3-c\end{pmatrix}\qquad R(0.2,0.2)=\begin{pmatrix} 0.5&0.2\\0.2&1.1\end{pmatrix}.
\end{equation}\noindent
There are two parameters. Recall that the parameter $a=r_{v,f}$ denotes the number of fit persons infected by an infectious vulnerable person; the parameter $c=r_{f,v}$ the number of vulnerables infected by an infectious fit person. The parameter $c$ varies over $[0,0.8]$ for six values of $a$, $a=0,0.1,\ldots,0.5$ in Figure~\ref{fepi1}, left. This figure and Figure~\ref{fepi2} show plots of the number of deaths for the whole population (solid curves) and for the vulnerable (dotted curves). The solid and dotted horizontal lines indicate the number of deaths in the corresponding homogeneous models with $r_0=1.1$ (Figure~\ref{fepi1}, left), $r_0=1$ (Figure~\ref{fepi2}, left), and $r_0=0.95$ (Figure~\ref{fepi2},right). For $r_0=1.1$ the homogeneous model yields a mortality of 2123.9 of which the majority, 5/6, is vulnerable. The mortality for the homogeneous model with $r_0=1$ is 168.7 and for $r_0=0.95$ it is 23.4. This paper focuses on the supercritical case $r_0=1.1$. 

The deaths are obtained from the infections as binomial variables:
\begin{verbatim}
   d<-c(rbinom(1,j[1],0.01), rbinom(1,j[2],0.001));
\end{verbatim}
For the left side of Figure~\ref{fepi1} the mean is computed over a hundred thousand simulations for the matrices $R(a,c)$, with $a$ in $0:5/10$ and $c$ in $0:40/50$.

The plot for $r_v=0.8$ and $r_f=1.1$ with $r_0=1$, Figure~\ref{fepi2}, left, paints a darker picture. For $a=0.2$ (green curve) and $c=0.1$ the overall death toll is twice as high as in the homogeneous case. Mortality among the fit is one third of the total rather than one sixth as in the homogeneous case. The reason for this behaviour? In the heterogeneous case there will be a full blown epidemic among the fit because $r_f=1.05>1$. The constant $c=r_{f,v}>0$ will pull the vulnerable into the epidemic. The overall reproduction number here has the critical value $r_0=1$. In the homogeneous model epidemics will be short and die out. 

The right side of Figure~\ref{fepi2} shows the subcritical case, $r_0=0.95$. The adverse effects of shielding the vulnerable here are already apparent for $c=0$. The mortality curves are decreasing. The excess death toll is small, less than ten. It is of little interest since it reflects the size of the contiguous initial infection rather than of the population.

In all three figures the situation brightens for $c$ in the upper half of the interval $[0,0.8]$. The total mortality decreases dramatically as $c\to0.8$.  The low mortality for $R(0.2,0.8)$ on first sight is a mystery. A large value $r_{f,v}=0.8$ indicates many infections from fit to vulnerable. Why should the mortality almost vanish? Restrict attention to the fit. First assume $a=r_{v,f}=0$. The top row of $R$ then is $(0.7,0)$, but if we turn to the fit we see that the infections will die out since $0.5<1$. There also are infections from the fit to the vulnerable, but these do not concern the fit since we have assumed that the vulnerable do not infect the fit, $a=r_{v,f}=0$. The infections among the vulnerable will also die out (since $r_{v,v}=0.7<1$), apart from the import from the fit, but the import will die down as the epidemic among the fit dies out. If $a=r_{v,f}$ is positive this will not alter the situation as long as $a$ is small. Section~\ref{sa1} contains a more detailed analysis.

\begin{figure}[htp]
\centering
\includegraphics[width=0.48\linewidth]{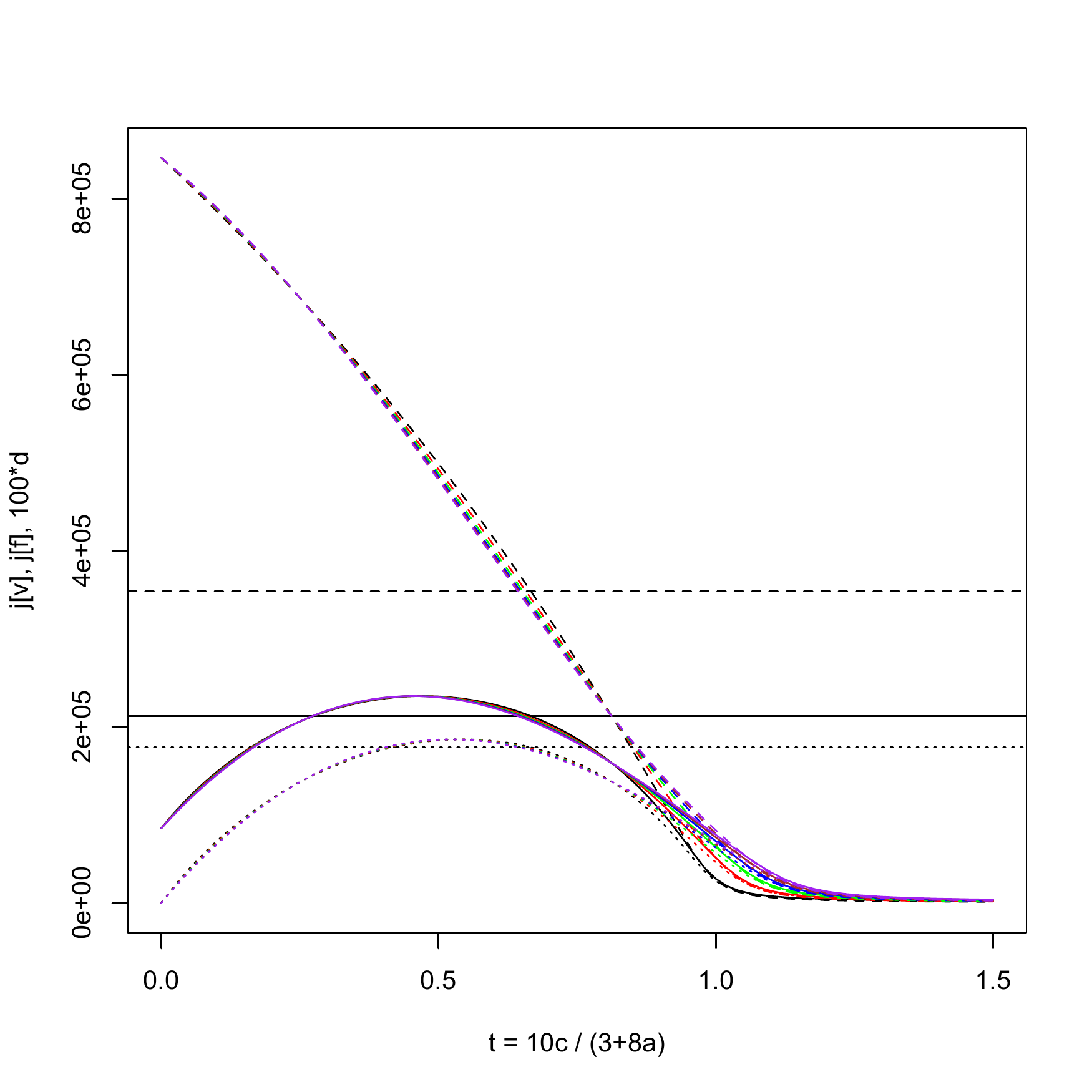}
\hspace{0pt}
\includegraphics[width=0.48\linewidth]{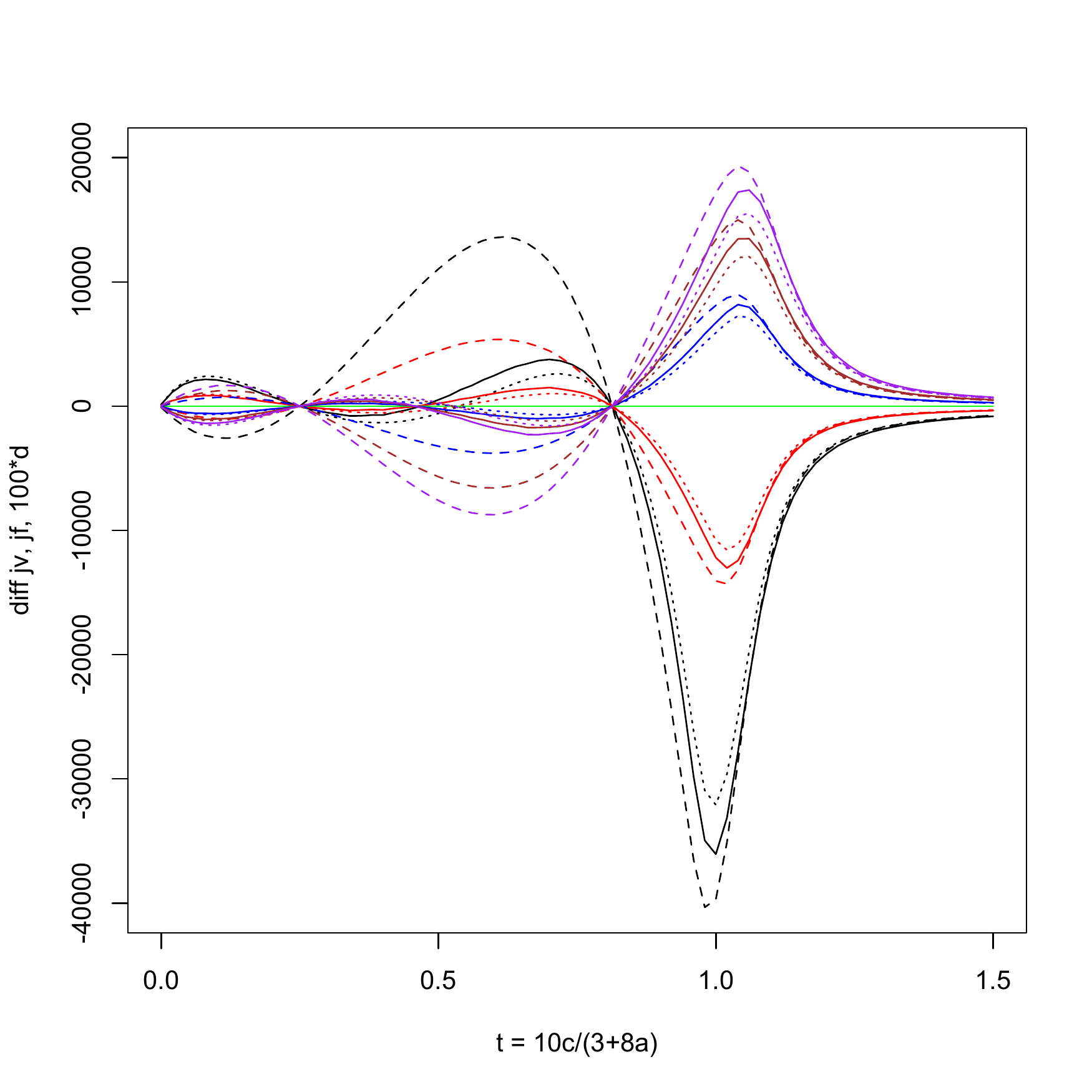}
\caption{On the left $j_a[v](t)$ (dotted) and $j_a[f](t)$ (dashed), the number of infections among the vulnerable and the fit, and $100d_a(t)$ (solid), the number of deaths, for $R(a,(0.3+0.8a)*t)$. On the right the differences with the standard graph, $a=0.2$.} 
\label{ftac} 
\end{figure}

The six mortality curves in the left part of Figure~\ref{fepi1} all have the same shape and the same maximum. (So too for the six curves in the subfigures in Figure~\ref{fepi2}.) This suggests that the six curves can be derived from a common curve by suitable affine transformations of the horizontal coordinate. Introduce a new variable, $t$, and write $c=(0.3+0.8a)*t$. The left side of Figure~\ref{ftac} plots $j_a[v](t)$ (dotted) and $j_a[f](t)$ (dashed), the number of infections among the vulnerable and the fit, and a multiple of the death toll, $100d_a(t)$ (solid) for the two type binomial model with reproduction matrix
\begin{equation}\label{qRt}
R_a(t)=R(a,c)\qquad0\le t\le 1.5, \quad a=0,0.1,\ldots,0.5;\qquad c=(0.3+0.8a)*t
\end{equation}\noindent 
in the same colours as the Figures~\ref{fepi1} and~\ref{fepi2}. The curves almost coincide. They all seem to vanish for $t\to1.5$. The role of the simple affine transformation $3+8a$ is not clear.

\section{Discussion}\label{s8}

One cannot argue with the result of a computation. One can argue about the interpretation. We restrict the discussion to two topics.

\noindent1) How realistic are our assumptions on the entries of the reproduction matrix?

In the basic model the reproduction matrix is
$$R=R(a,c)=\begin{pmatrix} r_{v,v}&r_{v,f}\\r_{f,v}&r_{f,f}\end{pmatrix}=\begin{pmatrix} 0.7-a&a\\c&1.3-c\end{pmatrix}.$$
A vulnerable infectious person infects on average 0.7 persons of whom $a$ are fit; a fit infectious person infects on average 1.3 persons of whom $c$ are vulnerable. Infection is due to contact. Social contacts of the healthy and young are more varied and more intense than for the old or sick. This difference has increased as the vulnerable have become more aware of their vulnerability. Human beings are social animals, but older people are perhaps better able to endure solitude and live with their thoughts and memories than the young. A factor $1.3/0.7\approx2$ may be excessive. In the example for the critical case, $r_0=1$, the factor is less, $1.05/0.9$. The effect is similar.

In mathematics it is good practice to vary one variable at a time. The effect depends on the variables which are kept constant. If one entry of the reproduction matrix goes down and the other three are constant the epidemic will be less severe. In a partition the lockdown for the vulnerable becomes stricter while at the same time it is relaxed for the fit. This approach makes it natural to assume $r_f$ and $r_v$ to be constant. We then compare the results of the epidemic for various values of $r_{f,v}$ and $r_{v,f}$. The homogeneous population with reproduction number $r_0=(r_v+2r_f)/3$ is the benchmark. This value of $r_0$ is somewhat arbitrary. It is not the reproduction number of the two type model. That depends on the proportion between the vulnerable and the fit among the infected. Figure~\ref{fcurve} shows that this proportion varies in the course of the epidemic.

The standard value $r_{v,f}=a=0.2$ is on the low side, but the graphs of the mortality for $a=0.0, 0.1, 0.2, 0.3, 0.4, 0.5$ all have the same shape. The assumption that one is free to vary $c=r_{f,v}$ is more problematic. The relation between the cross infection constants $a=r_{v,f}$ and $c=r_{f,v}$ is obscure. Contact is symmetric. Person A is within two meters of person B if and only if person B is within two meters of person A. If one replaces the infection rates $r_{v,f}$ and $r_{f,v}$ by contact rates $c_{v,f}$ and $c_{f,v}$ then symmetry of contact implies the conservation law: $n_fc_{f,v}=n_vc_{v,f}$. Contact is symmetric, but its effect on infection not. A healthy young person will cough with more force than a feeble old person. This may make the fit more infectious than the vulnerable. Thus there are indications that adults infect children but children hardly infect adults~\cite{H20, M20, K20}. (There also is contrary evidence~\cite{L20}.) There are more reasons for a lack of symmetry. If the vulnerable are tested at regular intervals and visits are only allowed when the test result is negative this will not affect $r_{f,v}$, but it will reduce the value of $r_{v,f}$ and hence increase the parameter $t$ in Figure~\ref{ftac}. 

A representative list of pairs, infector and infectee, together with age and medical condition, might help to determine the role of the values of $t>0.5$ in figure~\ref{ftac}.

\noindent2) Do the results apply to real life? 

The proportion of vulnerable to fit is 1:2. This is realistic for the Netherlands and perhaps for Italy and Japan. In countries with a younger population a proportion of 1:3 might be more appropriate. The population size is immaterial. The number of initial cases $(100,200)$ reflects the proportion of the vulnerable to the fit and is chosen large in order that the simulations all show full blown epidemics when the reproduction number is 1.05 or higher. The IFR of 0.004 is on the low side. It does not affect the results of the paper. The factor 10 between the IFR of 0.01 for the vulnerable and 0.001 for the fit is on the high side. The reader is invited to investigate the effect of increasing the IFR for the fit to 0.004.

What happens if one replaces the Reed-Frost model by a more realistic model? The assumption in the homogeneous model that everyone has the same probability of being infected is not realistic. In the two type model there is a reproduction matrix with four entries. If the more realistic model allows one to specify such a matrix, then -- as in the Reed-Frost model -- the maximal eigenvalue and the corresponding left eigenvector will determine the severity of the epidemic and the proportion of vulnerable to the fit among the infected. In the more realistic model the infection curves in Figure~\ref{fcurve} will have different shapes, but their relation to the dotted lines associated with the eigenvectors will be the same; the  positions of the points describing the total number infected will change, but the shape of the curve which they form might well again be a question mark. This is speculation. Whether the suppositions hold can only be determined by doing the necessary simulations and observations.

\section{Conclusion}\label{s9}

The results presented in this article are indicative rather than descriptive. They suggest that our intuition fails us in understanding how the parameters of the reproduction matrix affect the outcome of an epidemic in a population divided into two classes, the vulnerable and the fit.  It only takes a little effort to run the Reed-Frost model on a pc and simulate two type epidemics. Speed makes this model a viable alternative to our intuition. It may be a better guide to reality.

\section*{Acknowledgements}

The author thanks Alex van den Brandhof for his help in making the article accessible to the general reader.

\section*{Supplementary material}

\section{A non-technical explanation of the decrease in mortality}\label{sa1}

The conclusion that the mortality may be reduced by increasing the transmission of Covid-19 from the fit to the vulnerable is hard to swallow. For a mathematician the eigenvalue argument may be convincing. For computing the eigenvalues and eigenvectors high school mathematics suffices. One has to solve a quadratic equation and a system of two linear equations. These calculations will not convince the general reader. Here is a more intuitive argument:

Consider a billowing cloud of viruses in the region of the fit, a cloud which doubles in size every few days. In the region of the vulnerable the cloud of viruses shrinks and will fade away. Now assume the excess of viruses produced in the region of the fit every day is diverted to a region which is less hospitable and where the viruses will die out. The cloud in the region of the fit no longer grows. If one increases the part which is diverted by a fraction the cloud of viruses in the region of the fit will shrink at an exponential rate. This will be the case even if some of the viruses manage to find their way back to the region of the fit. 

Can one arrange things such that in the region of the vulnerable the cloud will fade away in spite of the influx from the region of the fit? One may argue that the influx is only temporary since the cloud in the region of the fit will fade away at an exponential rate, and so will the fraction which is diverted to the region of the susceptible. The situation becomes less clear if a fraction of the cloud above the region of the susceptible manages to return to the region of the fit. In order to handle the situation where there is traffic between the two regions in both directions we have to be more specific. It helps to look at steady states. We give two examples. In both cases the top row of the matrix $R$ is $(0.5, 0.2)$. The bottom row is $(c,1.3-c)$.

\begin{figure}[htp]
\centering
\includegraphics[width=0.48\linewidth]{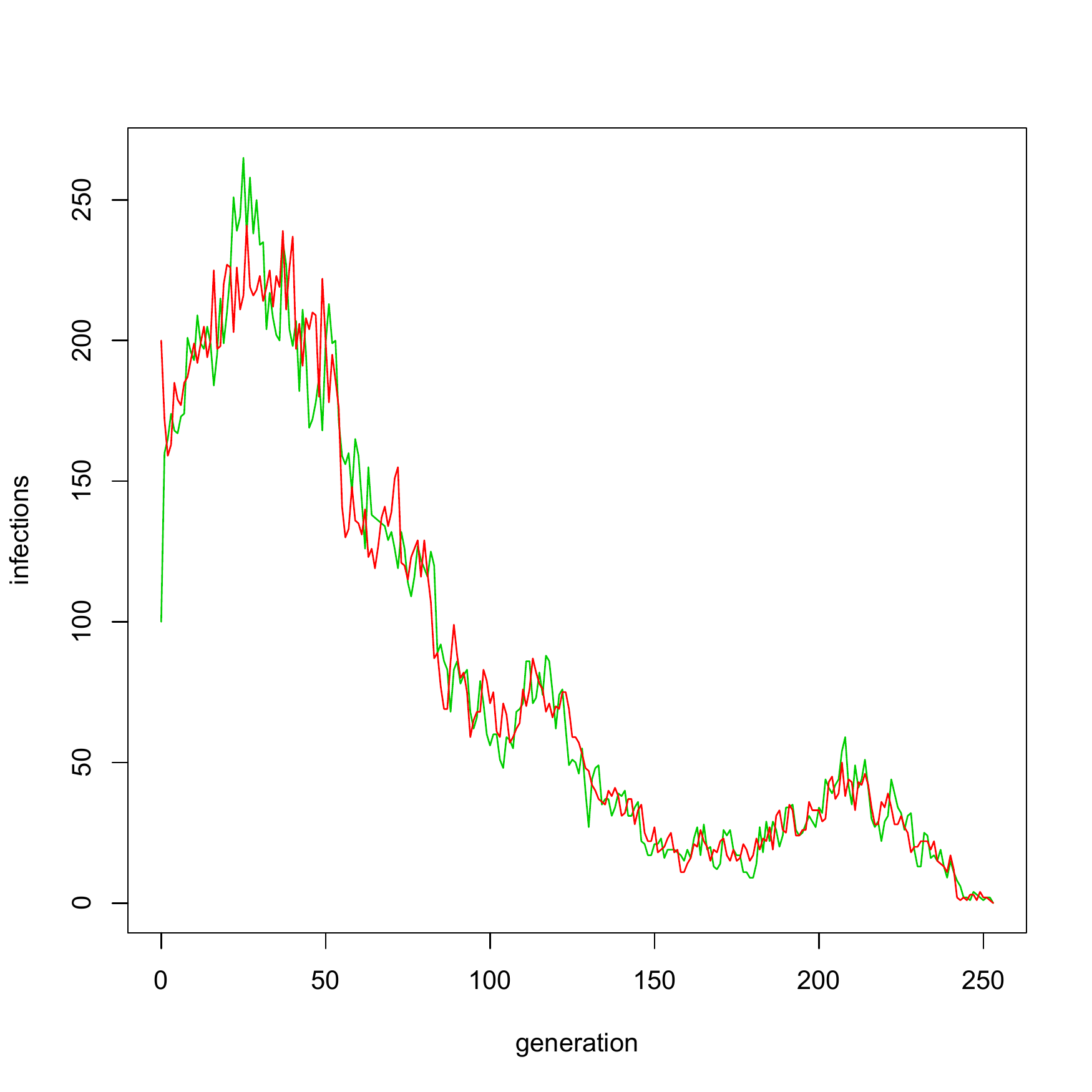}
\hspace{0pt}
\includegraphics[width=0.48\linewidth]{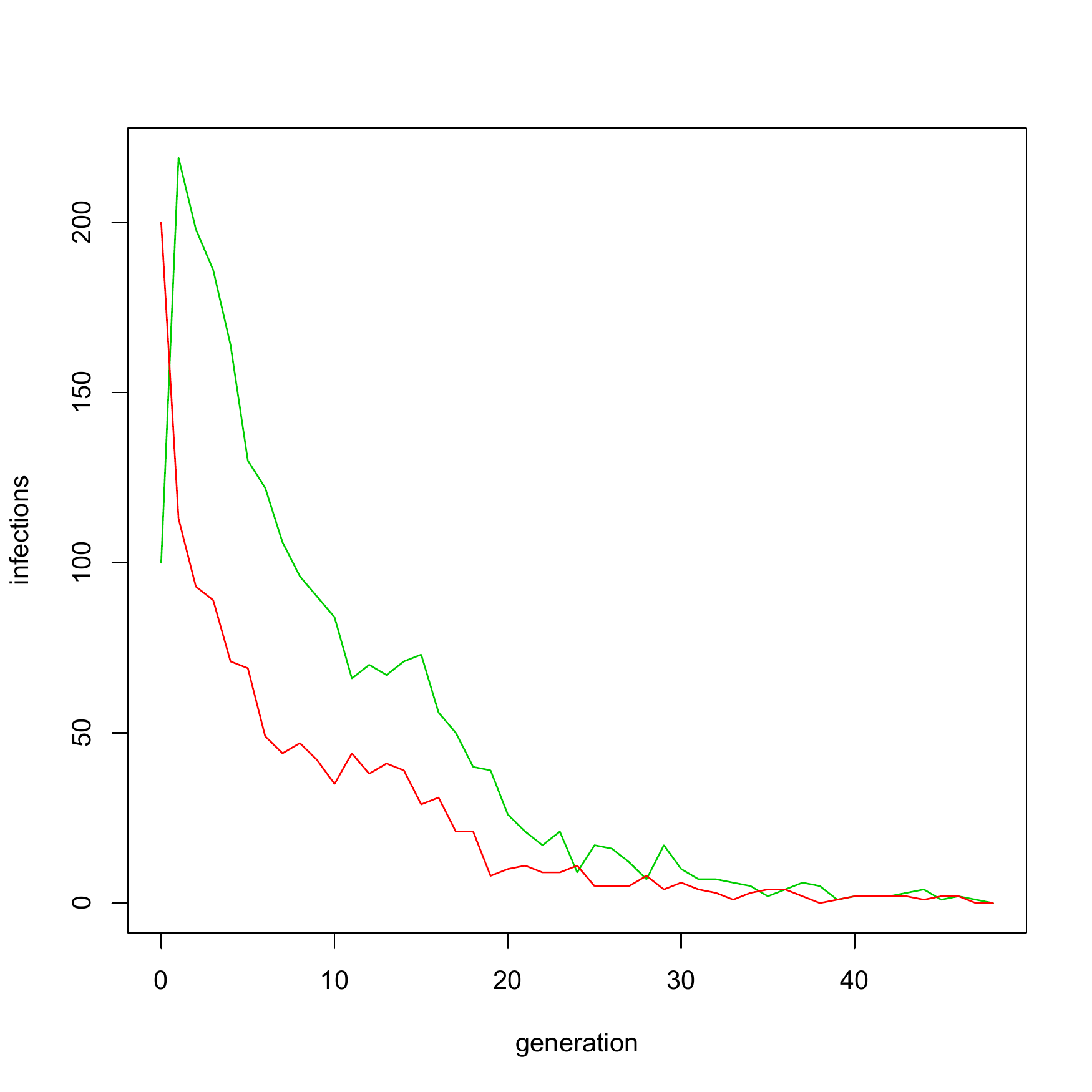}
\caption{Realizations of the epidemic $i_k(v)$ (green) and $i_k(f)$ (red) for $r_{v,f}=0.2$ and $r_{f,v}=0.5$ with a mean of 214.8 + 21.8 deaths on the left and $r_{f,v}=0.8$ with a mean of 23.1 + 1.3 deaths on the right. } 
\label{f258} 
\end{figure}

\begin{example}\rm
First assume $c=0.5$. Consider a situation where a thousand fit are infectious and a thousand vulnerable. The thousand vulnerable infect 500 vulnerable and 200 fit; the thousand fit infect $c*1000=500$ vulnerable and $(1.3-c)*1000=800$ fit. The result is a thousand new infections among the vulnerable and a thousand among the fit. At first sight nothing changes. But since there is a growing number of immune persons the new number of infectious persons decreases at an ever accelerating rate and the epidemic will die out. \hfill\mbox{$\lozenge$}
\end{example}

\begin{example}\rm
Take $c=0.8$. Let there be two thousand infectious vulnerable and a thousand fit. A calculation as above shows that these infect one thousand eight hundred vulnerable and nine hundred fit. The proportion $2:1$ between the vulnerable and the fit is preserved. In this case the epidemic dies out at an exponential rate.\hfill\mbox{$\lozenge$}
\end{example}

The pairs $(1000,1000)$ and $(2000,1000)$ are called eigenvectors of the corresponding matrices $R(0.2,0.5)$ and $R(0.2,0.8)$ and the factors $1.0$ and $0.9$ linking the number of new infections to the old are called the eigenvalues. Figure~\ref{f258} shows a realisation of each of these two type binomial epidemics. The proportions $(1:1)$ and $(2:1)$ are clearly visible in the number of infections of the vulnerable and the fit. 

\section{Eigenvalues and eigenvectors of the reproduction matrix}

Replace the binomial variables in~(\ref{qP}) by their expectations to obtain the deterministic mean process with
\begin{equation}\label{qri}
(i[1],i[2])=(n[1]*(1-q_1), n[2]*(1-q_2)).
\end{equation}\noindent
The sequence of successive infections $\vec i_k=(i_k[v],i_k[f])$ form the \emph{infection curve}.

The entries $1-Q[i,j]=R_{i,j}/n_0[j]$ in~(\ref{qmQ}) are of the order of $10^{-6}$. This suggests the approximation $(1-(1-p)^i)\approx ip$ and hence the recursion
\begin{equation}\label{qri0}
\vec i_{k+1}=\vec i_k*R*D_{k+1}\qquad D_{k+1}={\rm{diag}}(\vec n_{k+1}/\vec n_0)\qquad \vec n_{k+1}=\vec n_k-\vec i_k.
\end{equation}\noindent
The new value $\vec i_{k+1}$ of the vector $\vec i$ is written as a linear transformation $R_{k+1}$ of the old value $\vec i_k=(i_k[v],i_k[f])$, where $R_{k+1}=RD_{k+1}$ is a modulation of the reproduction matrix. 

The key to linear dynamical systems $\vec x_{k+1}=\vec x_k R$ is the left eigenvector associated with the largest eigenvalue.  In the multitype epidemic the maximal eigenvalue of the reproduction matrix $R$ determines the severity of the epidemic; the corresponding left eigenvector determines the proportion of vulnerable to fit among the infected. 

These words should not be taken literally. The relation between the eigenvalues and eigenvectors of the reproduction matrix $R$ and the course of the epidemic is not perfect as one sees on comparing the curves on the left side and the right side in Figure~\ref{fepi1}. The coordinates associated with the left eigenvectors of $R$ make $R$ diagonal but destroy the diagonality of the modulator $D_{k+1}$ in the recursion~(\ref{qri0}).

\begin{figure}[htp]
\centering
\includegraphics[width=1.0\linewidth]{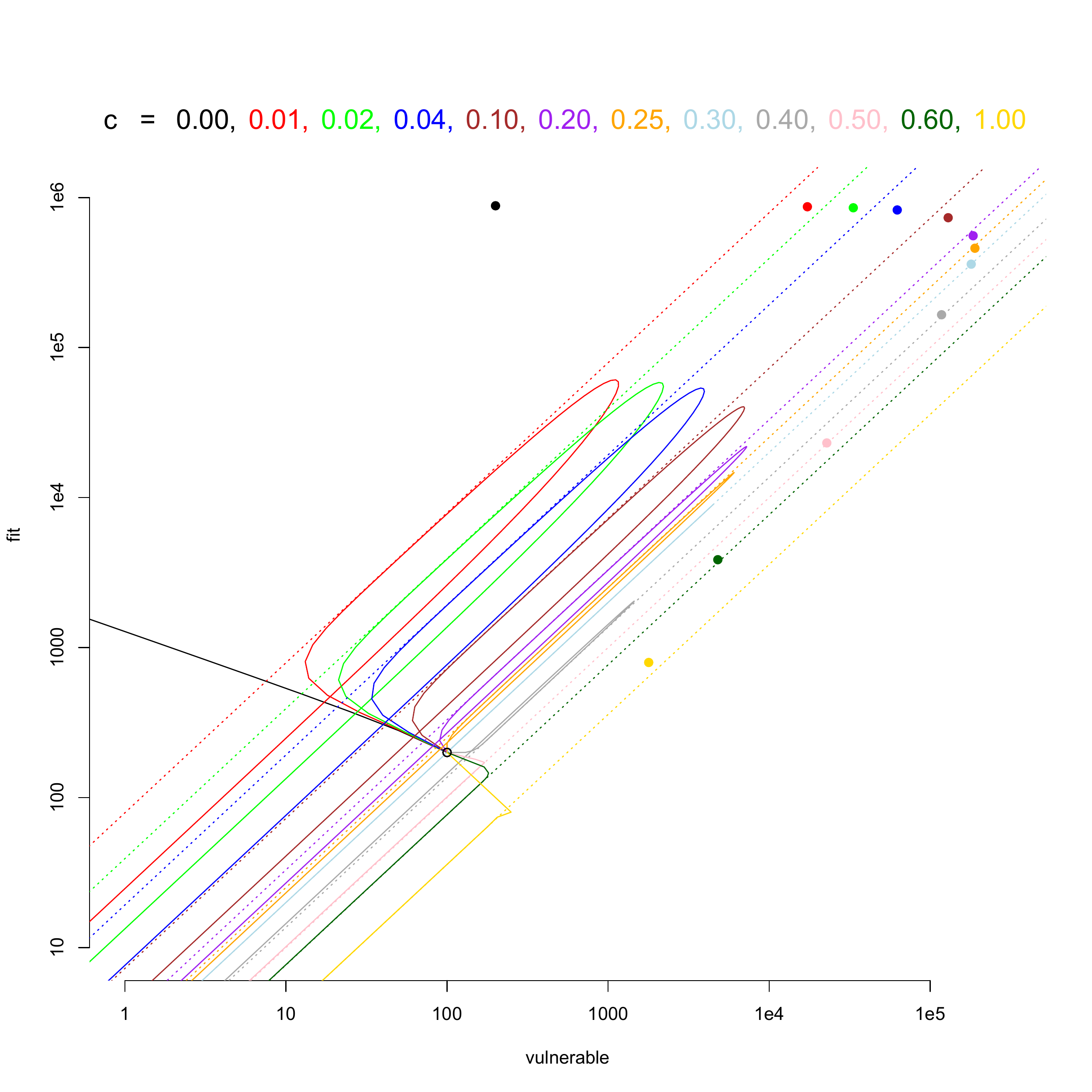}
\caption{Infection curves for $R(0.2,c)$ for various values of $c$, and the total number of infections, $(j_c[v],j_c[f])$.} 
\label{fcurve} 
\end{figure}

Figure~\ref{fcurve} attempts to give an impression of the changes in the epidemic as the parameter $c$ in the matrix $R(0.2,c)$ in~(\ref{qR}) varies from 0 to 1 over the twelve values listed above the plot. For each $c$ the infection curve starts at $\vec i_0=(100,200)$ and finally moves off to $(0,0)$ after attaining values which may be in excess of ten thousand, and the dot $\vec j_c=(j_c[v],j_c[f])$ depicts the total number of infections. The logarithmic scale on the two axes allows us to compress all this information into one figure. The logarithmic scale transforms rays $y=cx$ in the positive quadrant into lines $\log_{10}y=\log_{10}x +\log_{10}c$ parallel to the diagonal. Each dotted line corresponds to the ray through the left eigenvector $\vec e_c$ associated with the maximal eigenvalue of $R(0.2,c)$. Note that the quotient $i_k[v]/i_k[f]$ starts at $1/2$ for $k=0$, quickly approaches the quotient $e_c[v]/e_c[f]$ of the eigenvector, and then slowly backs off towards $1/2$. As a result the quotient $j_c[v]/j_c[f]$ lies between $e_c[v]/e_c[f]$ and $1/2$. Only the light blue infection curve (for $c=0.3$) lies on the line associated with the eigenvector and so does the light blue point depicting the total number of infections. That is because the vector $(1,2)$ is a left eigenvector of the matrix $R(0.2, c)$ for $c=0.3$. The coloured points form a question mark. As $c$ moves away from zero the total number of infections among the vulnerable, $j_c[v]$, increases rapidly, while the number among the fit decreases. The logarithmic scale makes it impossible to see what happens to the sum $j_c[v]+j_c[f]$. However beyond the orange point ($c=0.25$) both $j_c[v]$ and $j_c[f]$ decrease as $c$ increases. The number of infections among the vulnerable, $j_c[v]$ decreases from slightly more than a hundred thousand to slightly more than a thousand. A policy directed at shielding the vulnerable should try to \emph{maximize} the value of $c=r_{f,v}$ for the given boundary conditions $r_f=1.3$, $r_v=0.7$, $r_{v,f}=0.2$. Lines parallel to the diagonal through the coloured points drift to the South East as $c$ increases. The proportion of deaths for the vulnerable and the fit, $d_c[v]/d_c[f]=10j_c[v]/j_c[f]$, increases steadily from 0.002 to 22 as $c$ moves from 0 to 1. This may be expressed in the oracular rule: 
\centerline{``Sacrifice the vulnerable to save the vulnerable.''}

For the homogeneous Reed-Frost model there exists~\cite{B10} a simple and intuitive good approximation $\tau=\tau(r_0)$ to the fraction $j/n_0$ of infected when the reproduction number $r_0$ exceeds 1. It solves the equation $1-\tau=e^{-r_0\tau}$. Such a simple expression for the number infected is not available for multitype models. One may use $\tau=\tau(\rho)$, where $\rho$ is the largest eigenvalue of the reproduction matrix $R$, to estimate the total number of infections and use the corresponding left eigenvector to divide these among the vulnerable and the fit:
\begin{equation}\label{qej}
(j_v,j_f)=\vec i_0+\tau(\rho)*(n_0[v]-i_0[v]+n_0[f]-i_0[f])*\vec e/(e[1]+e[2]).
\end{equation}\noindent
This ad hoc estimate is not very accurate. 

Our benchmark is the mean value of $(j_v,j_f)$ for a hundred thousand simulations of the stochastic Reed-Frost model (RF). The table below lists estimates of the death toll for the two type model with reproduction matrix $R(0.2,c)$ for various values of $c$. It gives an indication of the accuracy of the three models E, L, D defined by~(\ref{qej}), (\ref{qri0}) and (\ref{qri}) compared to the stochastic model RF in~(\ref{qP}).  

\begin{center}\begin{tabular}{crrr rrrr rrrr}
$c$ =&0.0&0.1&0.2&0.3&0.4&0.5&0.6&0.7&0.8&0.9&1.0\cr
E&1270&2169&2443&2115&1260&&&&&&\cr 
L&884&2027&2404&2158&1342&251&52&31&24&21&19\cr 
D&848&1963&2347&2124&1332&251&52&31&24&21&19\cr 
RF&848&1963&2347&2124&1329&237&52&32&24&21&19
\end{tabular}\end{center}
\centerline{The death toll for $R(0.2,c)$ in the Reed-Frost model and three approximations}

\end{document}